\documentclass[francais]{gretsi}

\usepackage[utf8]{inputenc}

\usepackage[T1]{fontenc}
\usepackage{hyperref}
\usepackage{amsmath, amssymb, amsfonts}
\usepackage{subcaption}
\usepackage{xcolor}

\begin{document}

\newcommand{\modified}[1]{{\color{red}#1}} 


\titre{Stéréophotométrie par \textit{Gaussian Splatting} et rendu inverse}

\auteurs{
  \auteur{Matéo}{Ducastel}{mateo.ducastel@unicaen.fr}{}
  \auteur{Yvain}{Quéau}{yvain.queau@unicaen.fr}{}
  \auteur{David}{Tschumperlé}{david.tschumperle@unicaen.fr}{}
}

\affils{
  \affil{}{Université Caen Normandie, ENSICAEN, CNRS, Normandie Univ, GREYC UMR 6072, F-14000 Caen, France
  }
}

\resume{Les algorithmes récents de l'état de l'art en stéréophotométrie recourent à des réseaux de neurones et fonctionnent soit par apprentissage à priori, soit par optimisation par rendu inverse. Ici, nous revisitons le problème de la stéréophotométrie calibrée en exploitant les avancées récentes en rendu inverse 3D, avec le formalisme des \textit{Gaussian Splatting}. Ceci nous permet de paramétriser la scène 3D à reconstruire, et de l'optimiser de manière plus explicable. Notre approche intègre un modèle simplifié de représentation de la lumière, et montre le potentiel du moteur de rendu des \textit{Gaussian Splatting} pour le problème de la stéréophotométrie.}

\abstract{Recent state-of-the-art algorithms in photometric stereo rely on neural networks and operate either through prior learning or inverse rendering optimization. Here, we revisit the problem of calibrated photometric stereo by leveraging recent advances in 3D inverse rendering using the \textit{Gaussian Splatting} formalism. This allows us to parameterize the 3D scene to be reconstructed and optimize it in a more interpretable manner. Our approach incorporates a simplified model for light representation and demonstrates the potential of the \textit{Gaussian Splatting} rendering engine for the photometric stereo problem.}

\maketitle


\section{Introduction}

L'un des objectifs de la stéréophotométrie est d'estimer une carte de normales 3D décrivant la surface d'un objet à partir de plusieurs photographies prises avec une caméra fixe, mais avec des positions d'éclairage différentes. 
Ici, nous proposons d'explorer une nouvelle approche pour résoudre le problème de la stéréophotométrie, approche fondée sur le formalisme récent des \textit{Gaussian Splatting}.
Ce formalisme consiste à modéliser une scène comme un champ de gaussiennes dans un espace 3D, auxquelles sont attachés différents attributs (orientation, taille et couleur notamment).
Un moteur de rendu dédié permet par la suite de régénérer rapidement un rendu de la scène depuis n'importe quel point de vue. 
Ce qui est remarquable, c'est que ce moteur est différentiable, et qu'il peut donc intervenir dans des méthodes d'optimisation classiques (de type descente de gradient) afin d'estimer les paramètres des gaussiennes de manière à ce que les rendus des scènes resynthétisées soient les plus proches possibles des images d'entrée (résolution d'un problème inverse).
Nous proposons donc de revisiter le problème de la stéréophotométrie à l'aune de ce formalisme. \\

\textbf{\emph{État de l'art.}}
L'un des atouts majeurs de la stéréophotométrie est qu'elle permet de reconstruire des détails fins des scènes imagées, avec par exemple des applications en numérisation 3D d'œuvres pour le patrimoine \cite{LAURENT202543} ou encore la détection de défauts sur des pièces usinées \cite{WANG2023e02561}.
L’approche historique en stéréophotométrie est celle de Woodham \cite{10.1117/12.7972479} qui définit un modèle pour estimer la normale en chaque point d’un objet à partir de plusieurs observations sous différentes illuminations. Ce modèle suppose que l’objet possède une réflectance lambertienne, son apparence ne dépendant donc que de l'angle entre normale et la direction de la lumière :
\begin{equation} \label{eq:1} I_{(x,y)} \propto l^T \cdot n_{(x,y)}, \end{equation}

où $I$ est l'image en niveaux de gris, $l \in \mathbb{S}^2$ la direction de la lumière et $n_{(x,y)} \in \mathbb{S}^2$ la normale de l'objet observé en chaque point $(x,y)$ de l'image. 
Ce modèle peut être inversé grâce à une optimisation par moindres carrés, avec cependant plusieurs limitations, notamment l’incapacité à traiter les ombres portées et les réflexions spéculaires, ce qui la rend en pratique peu robuste pour les matériaux et les scènes complexes.
Suite à cette première approche pionnière, la majorité des méthodes qui ont suivi ont considéré les ombres portées et les réflexions spéculaires comme des données aberrantes, tout en développant des approches d'optimisation plus complexes et robustes afin de stabiliser la convergence.

Plus récemment, des approches exploitant l’apprentissage automatique profond ont vu le jour pour surmonter ces limitations. Hardy et al. \cite{HARDY2024104093} ont proposé une méthode d’apprentissage supervisé multi-échelle dans laquelle un réseau de neurones est entraîné sur une base de données contenant une grande diversité de formes et de matériaux afin d’estimer directement la carte des normales. Yamaguchi et al. \cite{Yamaguchi_2025_WACV} ont utilisé une estimation de carte de profondeur obtenue via \textit{Depth-Anything} pour initialiser grossièrement la carte de normales avant de l'affiner grâce à un réseau de neurones dédié. Contrairement aux méthodes historiques, ces nouvelles approches ne cherchent pas à respecter un modèle physique explicite, mais s’adaptent automatiquement aux surfaces aux propriétés de réflexion complexes. Toutefois, leur performance dépend fortement de la qualité et de la diversité du jeu de données d’entraînement, ce qui peut limiter leur capacité de généralisation aux matériaux non vus lors de l’apprentissage. De plus, elles nécessitent des machines puissantes pour entraîner et utiliser des modèles neuronaux profonds et induisent une perte d'explicabilité.

D’autres travaux récents adoptent une approche de rendu inverse plus poussée, en intégrant des estimations de réflectance spéculaire. Par exemple, Li et al. \cite{li2023dani} utilisent plusieurs petits réseaux de neurones spécialisés pour estimer des paramètres intermédiaires, tels que la position de la lumière et la carte de profondeur. Leur approche repose sur un modèle de rendu inverse différentiable, permettant l’utilisation de schémas d'optimisation itératifs pour ajuster les poids du réseau et minimiser l’écart entre l’image reconstruite et l’image observée. Une grande partie de l'optimisation est cependant réservée à l'estimation des ombres portées via un lancer de rayon effectué depuis chaque pixel, ce qui augmente grandement le temps d'optimisation (en pratique plusieurs dizaines de minutes). Yang et al. \cite{YANG2024108138} proposent d'utiliser la carte de profondeur pour estimer les ombres projetées et s'en servir pour mieux estimer les zones non éclairées de l'objet. Cependant, l'utilisation de réseaux pour estimer une première carte de profondeur requiert toujours des capacités de calcul importantes et des bases d’entraînement conséquentes.

La formulation d'un problème de reconstruction via un rendu inverse est intéressante, car elle rend le modèle plus flexible et adaptable à une grande variété de matériaux, à condition que celui-ci soit suffisamment généraliste. De plus, elle améliore l'explicabilité des résultats en prenant en compte un modèle physique de formation de l'image pour reconstruire l'image finale. Dans cette perspective, explorer de nouvelles approches de paramétrisation de surface est une piste prometteuse pour assurer à la fois l'explicabilité du processus de reconstruction et obtenir de bonnes performances, comme nous l’aborderons dans la suite de cet article, en nous limitant à la réflectance lambertienne afin d'établir une preuve de concept.

\section{Formulation proposée}

\textbf{\emph{Gaussian Splatting.}}
L'objectif est donc de pouvoir retrouver des cartes de normales d'objets avec des matériaux lambertiens en inversant le modèle de formation de l'image. 
Nous nous intéressons à la méthode du \textit{Gaussian Splatting} développée par Kerbl et al. \cite{kerbl3Dgaussians} pour utiliser les avantages des méthodes de rendus inverse différentiables associés. Cette technique a été initialement développée pour la reconstruction 3D multi-vues.
Elle repose sur une représentation compacte et efficace de la géométrie et de l'apparence de la scène à l'aide de champs de gaussiennes placées dans un espace 3D.

Le \textit{Gaussian Splatting} utilise un ensemble de gaussiennes localisées dans l’espace, chacune étant définie par plusieurs paramètres fondamentaux (orientation, taille, couleur). Le rendu d’une scène modélisée de cette façon s’effectue par la suite en projetant ces gaussiennes en 2D sur l’image finale, selon un angle de vue défini par l'utilisateur. Chaque gaussienne est définie par un centre $p_k \in \mathbb{R}^3$ ainsi qu’une matrice de covariance $\Sigma$ de telle sorte que pour tout point $p \in \mathbb{R}^3$ :

\begin{equation} 
    \label{eq:gaussian} 
    \mathcal{G}_k(p) =  \exp\left (-\frac{1}{2}(p-p_k)^T\Sigma_k^{-1}(p-p_k)\right).
\end{equation}

La matrice de covariance $\Sigma = RSS^TRT$ est définie par une matrice de rotation $R$ et une matrice de mise à l'échelle $S$ afin d'assurer qu'elle soit semi-définie positive.

Pour effectuer le \textit{splatting} des gaussiennes en chaque pixel $x$ de l'image, il faut tout d'abord obtenir le point 3D à évaluer sur chaque gaussienne. Pour cela, on récupère les coordonnées 3D de la projection du centre $p_k$ vers le rayon $r_x(t) = o + td_x$ lancé depuis la caméra de position $o$ selon la direction de visée $d_x$. La position sur le rayon retenue est celle qui minimise la distance orthogonale au centre $p_k$, notée $t_{min}$ :

\begin{equation} 
    \label{eq:projection} 
    \widehat{\mathcal{G}}(x) = \mathcal{G}(r_x(t_{min}))
\end{equation}

Une couleur est également attribuée à chaque gaussienne et est définie à l’aide d’harmoniques sphériques, ce qui permet à la couleur perçue de varier dynamiquement en fonction de l’angle d’observation. Enfin, une opacité est associée à chaque gaussienne afin de moduler son influence dans le processus de rendu, en contrôlant son importance relative dans le mélange des contributions via un processus d'\textit{alpha blending} en prenant les $k$ gaussiennes d'avant en arrière jusqu'à atteindre une saturation de l'opacité à $\alpha = 1$. On obtient la couleur $c(x)$ du pixel $x$ par l'équation suivante :

\begin{equation} 
    \label{eq:color} 
    c(x) = \sum_{k=1}^K c_k \alpha_k \widehat{\mathcal{G}}_k(x) \prod_{j=1}^{k-1}(1-\alpha_j \widehat{\mathcal{G}}_j(x))
\end{equation}

où $\alpha_k$ est le paramètre d'opacité et $c_k$ la couleur selon le point de vue concerné.
Un des principaux avantages du \textit{Gaussian Splatting} est que l’ensemble du processus est différentiable presque partout, ce qui permet d’optimiser les paramètres des modèles gaussiens par descente de gradient. L’optimisation repose sur une fonction de perte $L_1$ mesurant la différence entre l’image rendue et une image de référence (vérité terrain). Cette propriété rend donc la méthode particulièrement adaptée aux approches fondées sur l’apprentissage automatique et l’optimisation des paramètres de la scène.

En plus du $\textit{splatting}$ réalisé pour les couleurs, on peut reproduire ces étapes pour obtenir une carte de profondeur. \`A partir de cette carte de profondeur, il est naturellement possible d'estimer une carte de normales en calculant son gradient. Cependant, celle-ci ne sera pas forcément liée à la géométrie de l'objet car la nature volumétrique des gaussiennes 3D est en conflit avec la représentation de surfaces.

Pour résoudre ce problème, Huang et al. \cite{Huang2DGS2024} proposent une extension du \textit{Gaussian Splatting} en contraignant la gaussienne sur un espace tangent paramétré tel que le point $P$ sur le plan de la gaussienne 2D de coordonnées $(u,v)$ a pour équation : 

\begin{equation} 
    \label{eq:point_2d_gaussian} 
    P(u,v) = p_k + s_ut_uu + s_vt_vv = H(u,v,1,1)^T
\end{equation}

\begin{equation} 
    \label{eq:matrix_2d} 
    \text{avec} \hspace{1pt} H = \begin{bmatrix}
    RS & p_k \\
    0 & 1 \\
  \end{bmatrix},
\end{equation}

où $p_k \in \mathbb{R}^3$ est toujours le centre de la gaussienne dans l'espace 3D, $R = [t_u,t_v,t_w]$ est la matrice de rotation définie par les deux vecteurs tangents $t_u$ et $t_v$ ainsi que le vecteur orthogonal $t_w = t_u \times t_v$ au plan tangent et $S = (s_u, s_v)$ est la matrice de mise à l’échelle où la dernière composante est 0 pour former une gaussienne 2D. Le point pris en compte est celui sur le plan de la gaussienne.
Grâce à cette contrainte, il devient possible d’obtenir une carte de normales en réalisant un \textit{splatting} des normales de la même manière que pour la couleur. 
Nous avons donc choisi de reprendre l'idée générale de ces travaux afin de l'adapter pour résoudre le problème de la stéréophotométrie, en modélisant et exploitant la carte de normales pour obtenir un rendu final.\\

En temps normal, l'initialisation du \textit{Gaussian Splatting} se fait par \textit{Structure From Motion} ce qui permet d'obtenir un nuage de points éparse grâce à plusieurs points de vue d'une même scène. Cependant, pour la stéréophotométrie un seul point de vue est disponible. Pour pallier à ce problème, nous plaçons donc les points sur un plan de coordonnées $z=0$ afin que le rendu soit similaire à celui voulu pour faciliter le début de l'optimisation. Une gaussienne est placée pour chaque pixel et la couleur associée à chaque gaussienne correspond au maximum de la couleur observée sur l'ensemble des images ${I_i}$ avec différents éclairages:

\begin{equation} 
    c(x,y) = \max \{ I_1(x,y), I_2(x,y), \dots, I_n(x,y) \}.
\end{equation}

Il est désormais possible de générer un rendu de la carte des normales ainsi que de l'albédo de l'objet. Dans le modèle de \textit{Gaussian Splatting}, les normales ne sont pas directement utilisées pour le rendu car la lumière n'est pas prise en compte, ce qui implique qu'un rendu depuis un point de vue fixe sera toujours identique. Pour prendre en compte les différences de rendu en fonction de la position de la lumière, nous utilisons donc le modèle de l'équation (\ref{eq:1}) en analysant si les surfaces sont alignées ou non avec la direction de la lumière pour calculer les ombres propres. \\

\textbf{\emph{Optimisation.}}
Une fois que le modèle a été choisi, il faut définir la méthode d'optimisation pour pouvoir modifier les paramètres des gaussiennes (position, matrice de covariance, couleur et opacité). Comme mentionné précédemment, le moteur de rendu utilise exclusivement des gaussiennes afin de représenter la scène, ce qui permet de calculer efficacement les gradients pour les optimiser par une descente de gradient. 

La fonction de perte $\mathcal{L}$ que nous avons utilisée se décompose en deux termes. Tout d'abord, un terme de perte photométrique $\mathcal{L}_c$ qui va mesurer l'écart entre le rendu obtenu et le rendu réel par une fonction de perte $L_1$. Ensuite, une régularisation de la normale $\mathcal{L}_n$ est ajoutée afin de s'assurer que les gaussiennes sont alignées avec la vraie surface de l'objet. Pour cela, on calcule une carte de normales avec le gradient de la carte de profondeur et on la compare avec la carte de normales obtenue via \textit{splatting}. On obtient la fonction de perte suivante :

\begin{equation} 
    \label{eq:loss} 
    \mathcal{L} = \mathcal{L}_c + \lambda \mathcal{L}_n.
\end{equation}

Le nombre de gaussiennes va aussi évoluer lors de l'optimisation. En particulier, les gaussiennes trop grandes ou dont l'opacité est trop faible vont être supprimées. Elles pourront être aussi clonées ou découpées lorsque plus de détails sont nécessaires. Les détails de ce procédé sont décrits dans \cite{kerbl3Dgaussians}. 

L'ensemble de cette méthode va donc nous permettre d'utiliser le \textit{Gaussian Splatting} pour la stéréophométrie calibrée.

\vspace{-0.75em}
\section{Résultats}

\begin{figure}[htb]
  \centering
  \includegraphics[width=1\columnwidth]{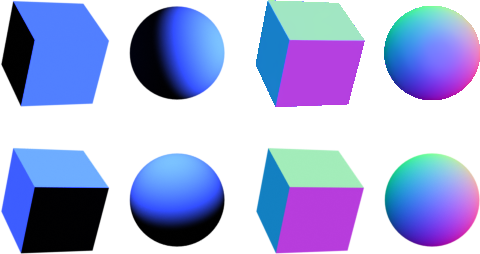}
  \caption{\small Exemples de rendus via Blender (à gauche haut/bas avec différentes positions de lumière). Normales estimées (après optimisation en haut à droite, et vérité terrain en bas à droite).}
  \label{results_blender}
\end{figure}

Notre procédé d'estimation de champs de normales par \emph{Gaussian Splatting} est testé sur différents objets.
Dans un premier temps, nous effectuons des tests sur des images synthétiques comprenant des formes géométriques simples, dont le rendu est effectué à l'aide de \emph{Blender} \cite{Blender}, un logiciel libre de modélisation et de rendu 3D.
Plusieurs images synthétiques sont ainsi générées, avec des positions de lumière différentes. Sur ces cas simples, notre algorithme retrouve des cartes de normales 3D cohérentes (Figure~\ref{results_blender}).



\begin{figure}[htb]
  \centering
    \includegraphics[width=1\columnwidth]{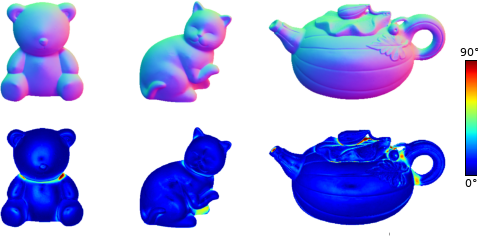}
  \caption{\small Normales estimées et carte d'erreur angulaire sur les objets \emph{Bear}, \emph{Cat}, et \emph{Pot1} de la base \emph{DiliGenT}. Les orientations des normales sont correctement estimées. Les zones d'erreurs d'estimation sont concentrées sur les zones d'ombres portées.}
  \label{results}
\end{figure}

Dans un second temps, nous nous comparons aux résultats des méthodes de l'état de l'art, grâce aux images du jeu de données
\emph{DiliGenT}, crée par Shi et al. \cite{7780772}, en particulier en estimant travaillant sur les objets \emph{Bear}, \emph{Cat} et \emph{Pot1}, qui sont des objets de la base avec des matériaux lambertiens
(Figure~\ref{results}).


\begin{figure}[htb]
  \centering
    \includegraphics[width=1\columnwidth]{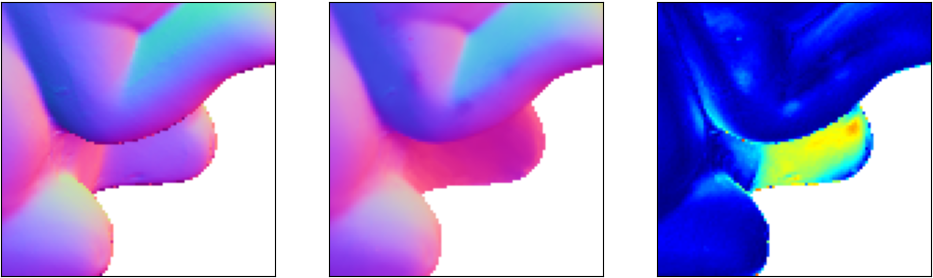}
  \caption{\small Vérité terrain des normales à gauche. Normales estimées au centre et carte d'erreur angulaire à droite. On voit que la partie cachée par des ombres portées a été mal estimée.}
  \label{diff}
\end{figure}

Remarquons que les zones où l'erreur angulaire est la plus élevée sont principalement les régions où une ombre portée est présente sur un grand nombre d'images d'entrée. Ceci s'explique par le fait que notre modèle ne prend pas encore en compte ce type d'ombre, qui sont les plus complexes à estimer à cause de leur nature globale. Nous illustrons ce phénomène en figure~\ref{diff}, où la patte arrière gauche du chat est cachée par l'ombre portée de la patte avant droite.

La métrique de qualité classique pour évaluer les modèles est la moyenne des erreurs angulaires (en degrés). Des comparaisons sur les trois objets sont présentées en Table~\ref{tab}, en les comparant à la méthode historique ainsi qu'à des méthodes plus récentes.

\begin{table}[htb]
    \begin{center}
    \begin{tabular}{l*{8}{c}}
        \toprule
          &   Cat & Bear & Pot1 \\
        \midrule
        Baseline \cite{10.1117/12.7972479} & 8.41 & 8.39 & 8.89 \\
        DANINet \cite{li2023dani} & 4.73 & 4.11 & 6.41 \\
        Uni MS-PS \cite{HARDY2024104093} & 3.45 & 3.14 & 4.12 \\
        GS-PS (Ours) & 8.26 & 8.13 & 10.11\\
        \bottomrule
    \end{tabular}
    \end{center}
    \caption{\small \label{tab}Performances sur le jeu de test DiliGenT, mesuré via l'erreur angulaire (en degrés).\vspace*{-1em}}
\end{table}

Qualitativement, nos premiers résultats sont proches de ceux obtenus par la méthode de Woodham \cite{10.1117/12.7972479}, ce qui est attendu, car nous considérons un modèle de formation d'image identique.
Néanmoins, revisiter ce problème d'estimation des champs de normales à l'aune des \emph{Gaussian Splatting} ouvre des perspectives prometteuses, notamment une possible gestion des matériaux spéculaires ainsi qu'une détection et une prise en compte possible des ombres portées dans les images d'entrées (voir section suivante).


Il faut cependant noter que d'autres paramètres peuvent impacter l'intérêt d'une méthode, par exemple l'explicabilité du modèle ou les ressources nécessaires pour le faire fonctionner. En particulier, notre méthode est plus explicable que les méthodes basées sur des réseaux de neurones profonds car elle optimise un modèle physique explicite de formation de l'image pour obtenir la carte de normales. Elle est également plus explicable que les méthodes de rendus inverse utilisant des petits réseaux car les paramètres que nous optimisons ont une interprétation géométrique, contrairement aux poids des réseaux de neurones.



\vspace*{-0.75em}
\section{Conclusions et travaux futurs}

Nos premiers résultats d'utilisation du \emph{Gaussian Splatting} pour résoudre le problème de l'estimation de champs de normales en stéréophotométrie sont très encourageants. Dans le futur, nous chercherons d'abord à affiner notre modèle afin de pouvoir gérer des matériaux non-lambertiens. Ceci est envisageable car il existe des moteurs de rendu spécifiques aux \emph{Gaussian Splatting} permettant de générer des réflexions spéculaires \cite{jiang2024gaussianshader,liang2024gs}.
De même, notre méthode n'estime actuellement que les ombres propres sur les surfaces des objets.
Or, il existe également des ombres portées qui apparaissent lorsqu'un point de la surface se trouve occulté par une autre zone de l'objet se trouvant sur la trajectoire de la lumière. Pour calculer la localisation de ces ombres portées, une technique classique est d'utiliser le \textit{Shadow Mapping}, introduit par Williams \cite{10.1145/965139.807402}.

\begin{figure}[htb]
  \centering
    \includegraphics[width=1\columnwidth]{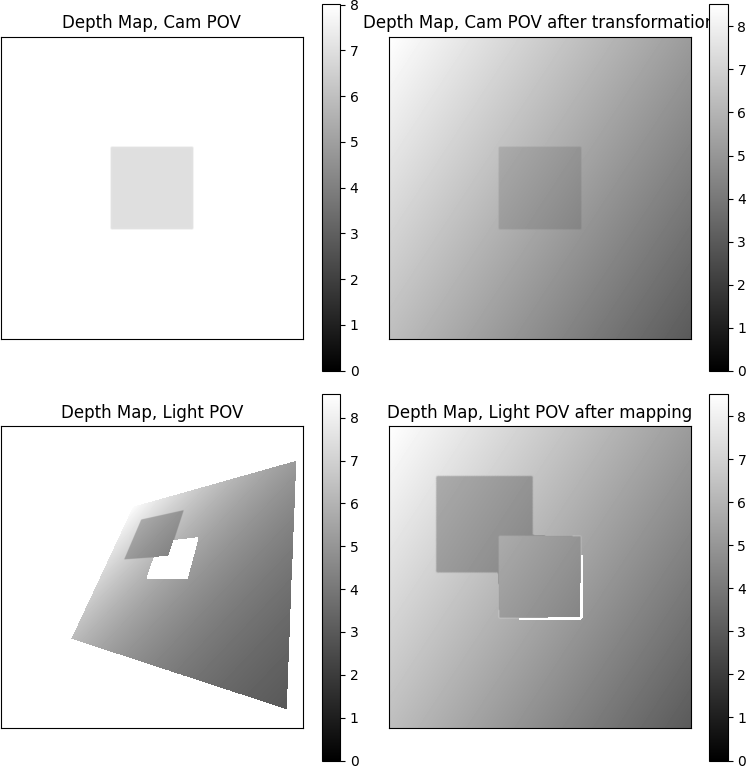}
  \caption{\small \'{E}tapes du \textit{Shadow Mapping} pour détecter les ombres portées. En haut à gauche, la carte de profondeur vue depuis la caméra. En haut à droite, la même carte après application d'une transformation rigide vers l'emplacement de la source lumière. En bas à gauche, la carte de profondeur du point de vue de la lumière. En bas à droite, la même carte après transformation vers l'espace de la caméra. Les deux images de droite sont comparées pour détecter les points correspondant aux ombres portées.}
  \label{shadow_mapping_steps}
\end{figure}

Le \emph{Shadow Mapping} consiste à effectuer un rendu de profondeur depuis le point de vue de la lumière, puis à comparer ce rendu avec les coordonnées des points 3D obtenus via la carte de profondeur générée depuis le point de vue de la caméra. Ce principe est illustré en Figure~\ref{shadow_mapping_steps} sur un objet synthétique simple (en utilisant le moteur de rendu du \textit{Gaussian Splatting}).

\small
\bibliography{biblio}


\end{document}